\documentclass[twocolumn,prb,byrevtex,floats,floatfix,showpacs]{revtex4}%
\usepackage{amsfonts}
\usepackage{graphicx}
\usepackage{amsmath}
\usepackage{amssymb}%
\providecommand{\U}[1]{\protect\rule{.1in}{.1in}}
\begin{document}
\title[Short title for running header]{Determination of the exchange anisotropy in perovskite antiferromagnets using
powder inelastic neutron scattering}
\author{R. J. McQueeney}
\affiliation{Department of Physics \& Astronomy and Ames Laboratory, Iowa State University, Ames, Iowa 50011}
\author{J.-Q. Yan}
\affiliation{Ames Laboratory, Iowa State University, Ames, Iowa 50011}
\author{S. Chang}
\altaffiliation{current address: National Institute of Standards and Technology, Center for Neutron Research, Gaithersburg, Maryland 20899}
\affiliation{Ames Laboratory, Iowa State University, Ames, Iowa 50011}
\author{J. Ma}
\affiliation{Department of Physics \& Astronomy, Iowa State University, Ames, Iowa 50011}


\begin{abstract}
A procedure is outlined for the determination of magnetic exchange constants
in anisotropic perovskite anitferromagnets using powder inelastic neutron
scattering. \ Spin wave densities-of-states are measured using time-of-flight
inelastic neutron scattering for LaMnO$_{3}$ ($A$-type antiferromagnet),
LaVO$_{3}$ ($C$-type), and LaFeO$_{3}$ ($G$-type) and compared to Heisenberg
model calculations. \ The anisotropy of in-plane ($J_{ab}$) and out-of-plane
($J_{c}$) exchange constants can be obtained from the data. \ The procedure
quickly determines the magnetic exchange interactions without the need for
single-crystal dispersion measurements and allows for rapid systematic studies
of the evolution of magnetism in perovskite systems.

\end{abstract}
\received[Version: June 20, 2008]{}
%{\today}

\maketitle

\section{Introduction}

Many important magnetic materials, such as colossal magnetoresistive
manganites and high temperature superconductors, are based on perovskite
transition metal oxides. The ground states and phases of these materials are
known to depend sensitively on the energy scales of magnetic, orbital,
vibrational, and electronic degrees-of-freedom.\cite{tokura2000, tokura2006}
Various spectroscopic techniques are employed to determine these energy
scales, and the coupling between them, in an effort to understand and control
the myriad of physical properties of these compounds.

The magnetic energy scale is set by the exchange energy between magnetic ions. It
is purely quantum mechanical in origin and can arise from many different
processes originating from the exchange of electrons between magnetic ions;
such as direct exchange, superexchange, and double exchange.\ In insulating
transition metal oxide materials, the superexchange interaction depends on the
overlap of metal $d$-orbitals on neighboring sites via oxygen ligands. The
rules for determining the sign (ferromagnetic or antiferromagnetic) and
strength of the superexchange between neighboring ions were established early
on.\cite{goodenough} A brief summary of these rules as applied to perovskites
is as follows; two half-filled (or empty) orbitals in a (180$^{o}$) bonding
configuration have antiferromagnetic (AF) exchange, while a half-filled and an
empty orbital in a bonding configuration have ferromagnetic (F) exchange.

In cubic perovskites, F exchange is not expected since all ionic sites are
equivalent. Such is the case for LaFeO$_{3}$, where each Fe$^{3+}$ ion has a
half-filled 3d$^{5}$configuration and all exchange interactions are AF,
leading to a $G$-type magnetic structure (with all neighboring magnetic ions aligned
antiparallel). \ However, metal ions with an orbital degeneracy are often relieved of this degeneracy by orbital ordering (due to Jahn-Teller distortions, for example). Orbital ordering can make neighboring ions inequivalent and often leads to the presence of both F and AF interactions in the nominally cubic perovskites. This is true for LaMnO$_{3}$, where a Jahn-Teller distortion elongates the oxygen octahedra and lifts the orbital degeneracy of the Mn$^{3+}$ $t_{2g}^{3}e_{g}^{1}$ ion. Staggering of the elongated axis in the $ab$-plane minimizes strain and causes ordering of $e_{g}(3x^{2}-r^{2}/3y^{2}-r^{2})$ orbitals. \ In the $ab$-plane, $e_{g}$
orbitals on all neighboring ions have a half-filled/empty configuration,
leading to F exchange, while overlaps of the $t_{2g}$ orbitals along the $c$-axis remain AF (half-filled/half-filled). \ The net result is the $A$-type magnetic structure
of ferromagnetic $ab$-planes coupled antiferromagnetically to neighboring planes along $c$.
\ In LaVO$_{3}$, the degeneracy of the V$^{3+}$ $t_{2g}^{2}$ ionic configuration is
lifted by orbital ordering resulting in full occupancy of the $xy$-orbital and
staggered occupancy of the $xz/yz$-orbitals in all three cubic directions.  This orbital ordering leads to the $C$-type magnetic structure of AF planes coupled ferromagnetically along $c$.

The energy of the exchange interactions in these and other perovskite magnets
are most commonly determined by measurement of the spin wave dispersions using
inelastic neutron scattering on single-crystal samples. Such measurements can
be time consuming (taking a week, or more, of measurement time) and depend on
the availability of large single-crystals. In this manuscript, we show that
the different exchange interactions in the $ab$-plane and along the $c$-axis (termed the magnetic exchange anisotropy) in simple systems can be obtained rapidly by measurement of inelastic neutron scattering from powders. Inelastic neutron scattering from powder samples gives information on the spin wave spectrum related to the spin wave density-of-states (SWDOS). \ Such measurements can take as little as a few hours to complete and allow for efficient systematic studies of the dependence of exchange interactions on temperature and composition.

We present powder inelastic neutron data for the aforementioned compounds; LaFeO$_{3}$, LaVO$_{3}$, and LaMnO$_{3}$, as representative of $G$-, $C$-, and $A$-type antiferromagnets, respectively.\ The data are compared to calculations of the spin wave spectra and their neutron scattering cross-sections using a Heisenberg model. The results show that the magnetic exchange anisotropy of F and AF interactions can be determined in a straightforward manner from powder data.

\section{Spin waves in cubic perovskites}

Spin waves in the cubic perovskite insulators can be described by the
Heisenberg Hamiltonian. In the case where cubic symmetry is broken by charge
or orbital orderings, exchange interactions can become anisotropic. For the
simplest kind of anisotropy, the exchange within the perovskite $ab$-plane
($J_{ab}$) and that along the $c$-axis ($J_{c}$) have different values, and
can even have different signs. The Heisenberg Hamiltonian becomes,
\begin{equation}
H=-J_{ab}%
%TCIMACRO{\dsum \limits_{<i,j>\parallel a,b}}%
%BeginExpansion
{\displaystyle\sum\limits_{<i,j>\parallel a,b}}
%EndExpansion
\mathbf{S}_{i}\cdot\mathbf{S}_{j}-J_{c}%
%TCIMACRO{\dsum \limits_{<i,j>\parallel c}}%
%BeginExpansion
{\displaystyle\sum\limits_{<i,j>\parallel c}}
%EndExpansion
\mathbf{S}_{i}\cdot\mathbf{S}_{j}-g\mu_{B}H_{a}%
%TCIMACRO{\dsum \limits_{i}}%
%BeginExpansion
{\displaystyle\sum\limits_{i}}
%EndExpansion
\sigma_{i}S_{i}
\end{equation}
where $\mathbf{S}_{i}$ is the spin vector on the $i$th site. The subscripts ${<i,j>\parallel a,b}$ ($c$) indicate that sums are restricted to nearest neighbor spins in the $ab$ plane (along the $c$ axis).
are restricted to nearest neighbor spins in the $ab$-plane and along the $c$-axis.
\ Exchange energies are defined such that a positive $J$ represents
ferromagnetic exchange. \ Uniaxial single-ion anisotropy is represented by an
anisotropy field $H_{a}$ that acts on spin $S_{i}$ and points along the local
spin direction (given by $\sigma_{i}=\pm1$).

In the following, we identify four different magnetic structures, one of which
is ferromagnetic and the other three are antiferromagnetic varieties. The
structures are differentiated by the signs of $J_{ab}$ and $J_{c}$.
\begin{eqnarray}
\mbox{F-type}: J_{ab}>0, J_{c}>0 \nonumber \\
\mbox{G-type}: J_{ab}<0, J_{c}<0 \nonumber \\
\mbox{C-type}: J_{ab}<0, J_{c}>0 \nonumber \\
\mbox{A-type}: J_{ab}>0, J_{c}<0
\end{eqnarray}

\begin{figure}
[ptb]
\includegraphics[
scale=0.6
]%
{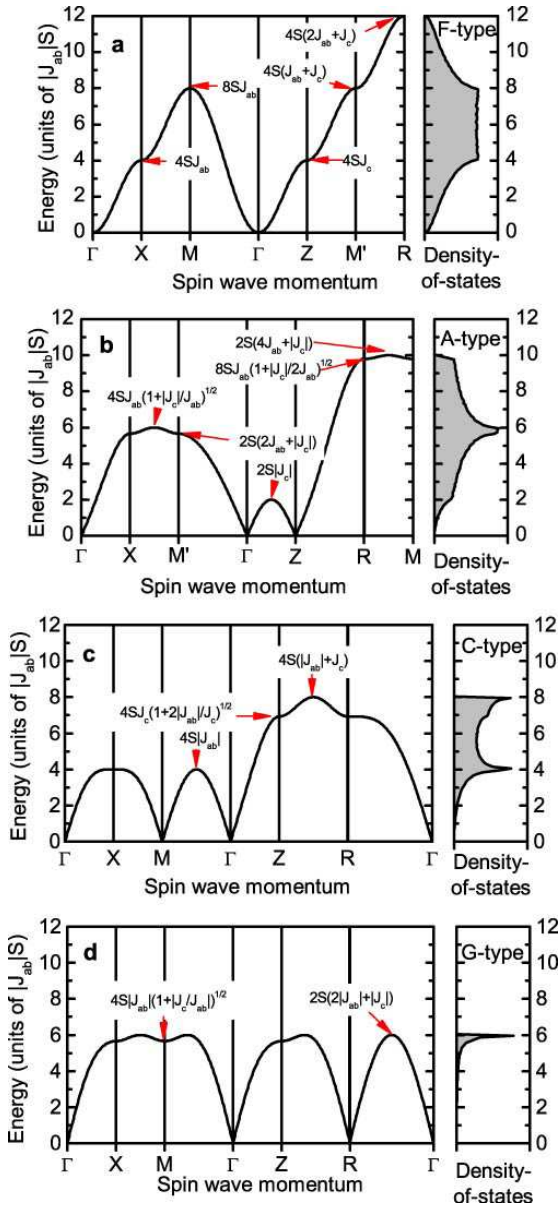}%
\caption{(color online) The spin wave dispersion along various symmetry
directions (left panels)\ and the spin wave density of states (right panel)
for (a)\ F-type, (b), A-type, (c), C-type, and (d) G-type perovskite
antiferromagnets. \ Red arrows and labels indicate the energies of the extrema in the dispersion that give rise to van Hove singularities in the density-of-states.}%
\label{fig_disp}%
\end{figure}

The spin wave dispersions for each type of magnetic ordering are obtained from a linear expansion (spin wave expansion) of the Heisenberg model.\cite{rackowski2002} When the single-ion anisotropy is zero, the dispersions are

\begin{eqnarray}
\hbar\omega_{F}\left(  \mathbf{q} \right)  & = & 2S\left[  2J_{ab}\left(  1-\gamma_{+}\left(  \mathbf{q}\right)  \right)  +J_{c}\left(  1-\gamma_{z}\left(\mathbf{q}\right)  \right)  \right]\nonumber\\
\hbar\omega_{G}\left(  \mathbf{q} \right)  & = & 2S\{\left(  2\left\vert J_{ab}\right\vert +\left\vert J_{c}\right\vert \right)  ^{2} \nonumber \\
 & & -\left[  2\left\vert J_{ab}\right\vert \gamma_{+}\left(  \mathbf{q}\right)  +\left\vert J_{c}\right\vert \gamma_{z}\left(  \mathbf{q}\right)  \right]  ^{2}\}^{1/2} \nonumber \\
\hbar\omega_{C}\left(  \mathbf{q} \right)  & = & 2S\{\left[  2\left\vert J_{ab} \right\vert +J_{c}\left(  1-\gamma_{z}\left(  \mathbf{q}\right)  \right) \right]^{2}\nonumber \\
& & -4J_{ab}^{2}\gamma_{+}^{2}\left( \mathbf{q}\right)  \}^{1/2}\nonumber\\
\hbar\omega_{A}\left(  \mathbf{q} \right)  & = & 2S\{\left[  2J_{ab}\left(  1-\gamma_{+}\left(  \mathbf{q}\right)  \right) \nonumber+\left\vert J_{c}\right\vert \right]^{2} \nonumber \\
& & -J_{c}^{2}\gamma_{z}^{2}\left(  \mathbf{q} \right)\}^{1/2}
\end{eqnarray}

where $\gamma_{+}\left(  \mathbf{q}\right)  =\frac{1}{2}\left(  \cos
q_{x}a+\cos q_{y}a\right)  $, $\gamma_{z}\left(  \mathbf{q}\right)  =\cos
q_{z}a$, $\mathbf{q}$ is the spin wave momentum, and $a$ is the cubic
perovskite lattice constant. The dispersions for each magnetic structure are
shown in fig. \ref{fig_disp} for the case where $\left\vert J_{ab}\right\vert
=|J_{c}|$. The notation for labeling the zone boundary reciprocal space
positions are found in Kovalev.\cite{kovalev} \ The spin wave
density-of-states (SWDOS) is the distribution of spin wave energies and is
determined by the summation over all wavevectors in the Brillouin zone
($\mathbf{q}$),
\begin{equation}
g\left(  \omega\right)  =%
%TCIMACRO{\dsum \limits_{\mathbf{q}}}%
%BeginExpansion
\frac{1}{N}{\displaystyle\sum\limits_{\mathbf{q}}}
%EndExpansion
\delta\left(  \omega-\omega\left(  \mathbf{q}\right)  \right)
\label{eqn1}
\end{equation}
The densities-of-states are also shown for the four magnetic structure types
in fig. \ref{fig_disp}. \ In addition, fig. \ref{fig_disp} indicates the
energies of the various extremal features in the SWDOS (van Hove singularities
(vHs)) for any $J_{ab}$ and $J_{c}$.

%TCIMACRO{\FRAME{ftbpFU}{3in}{2.0003in}{0pt}{\Qcb{(color online) The spin wave
%dispersion along various symmetry directions (right panels)\ and the spin wave
%density of states (left panel) for (a)\ F-type, (b), A-type, (c), C-type, and
%(d) G-type perovskite antiferromagnets. \ Energies of the extrema in the
%dispersion that give rise to Van Hove singularities in the density-of-states
%are indicated.}}{\Qlb{fig_disp}}{fig1_disp.eps}%
%maintain-aspect-ratio TRUE;  display "USEDEF";  valid_file "F";  width 3in;
%height 2.0003in;  depth 0pt;  original-width 0pt;  original-height 0pt;
%cropleft "0";  croptop "1";  cropright "1";  cropbottom "0";
%filename 'fig1_disp.eps';file-properties "XNPEU";}} }%
%BeginExpansion
%EndExpansion

\section{Spin wave density-of-states}

\begin{figure}
[ptb]
\begin{center}
\includegraphics[
]%
{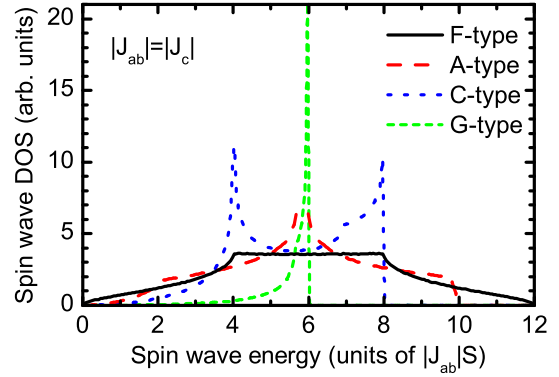}%
\caption{(color online)\ Spin wave density-of-states for F-, A-, C-, and
G-type antiferromagnets in the case where $|J_{ab}\left\vert =\right\vert
J_{c}|$.}%
\label{fig_dos1}%
\end{center}
\end{figure}

If $\left\vert J_{ab}\right\vert =\left\vert J_{c}\right\vert $, it is clear
that the four types of ordering are easily discernible from the spin wave
dispersions and densities-of-states. \ Figure \ref{fig_dos1} shows the SWDOS
for each type of ordering in the case of $\left\vert J_{ab}\right\vert
=\left\vert J_{c}\right\vert $. \ The maximum spin wave energy increases as
more ferromagnetic bonds are introduced. The F, A, C, and G-type structures
have the maximum spin wave energies of $12J_{ab}S$, $10J_{ab}S$, $8J_{ab}S$,
and $6J_{ab}S$, respectively.
%
%TCIMACRO{\FRAME{ftbpFU}{3.1825in}{2.4111in}{0pt}{\Qcb{(color online)\ Spin
%wave density-of-states for F-, A-, C-, and G-type antiferromagnets in the case
%where $|J_{ab}\left\vert =\right\vert J_{c}|$.}}{\Qlb{fig_dos1}}%
%{fig2.eps}{\special{ language "Scientific Word";  type "GRAPHIC";
%width 3.1825in;  height 2.4111in;  depth 0pt;  original-width 3.1384in;
%original-height 2.3705in;  cropleft "0";  croptop "1";  cropright "1";
%cropbottom "0";  filename 'fig2.eps';file-properties "XNPEU";}} }%
%BeginExpansion
%EndExpansion

\ In order to determine the exchange constants from powder samples, one must
consider the degree of information available in the SWDOS. When $\left\vert
J_{ab}\right\vert \neq\left\vert J_{c}\right\vert $, the positions of the vHs in the SWDOS allow identification of the exchange energies for each of the structure types:
%
%TCIMACRO{\FRAME{ftbpFU}{3.7291in}{3.5155in}{0pt}{\Qcb{(color online) Spin wave
%density-of-states for different ratios of the exchange $|J_{ab}/J_{c}|$ for
%(a) F-type, (b), A-type, (c) C-type, and (d) G-type antiferromagnets.}%
%}{\Qlb{fig_dos2}}{fig3.eps}{\special{ language "Scientific Word";
%type "GRAPHIC";  maintain-aspect-ratio TRUE;  display "USEDEF";
%valid_file "F";  width 3.7291in;  height 3.5155in;  depth 0pt;
%original-width 3.6815in;  original-height 3.4688in;  cropleft "0";
%filename 'fig3.eps';file-properties "XNPEU";}} }%
%BeginExpansion
\begin{figure}
[ptb]
\includegraphics[
scale=0.9
]%
{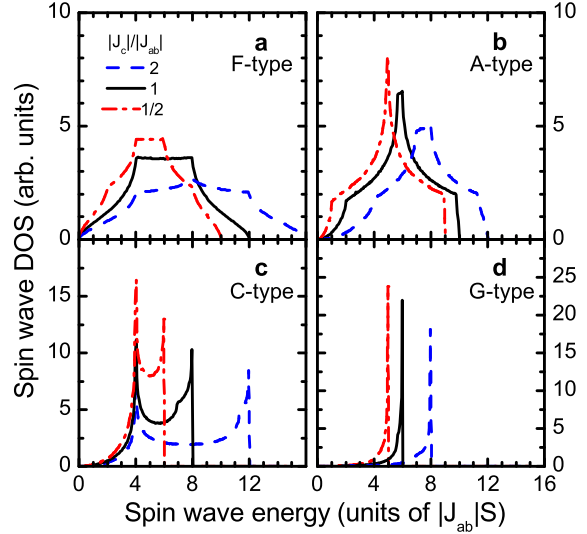}%
\caption{(color online) Spin wave density-of-states for different ratios of
the exchange $|J_{ab}/J_{c}|$ for (a) F-type, (b), A-type, (c) C-type, and (d)
G-type antiferromagnets.}%
\label{fig_dos2}%
\end{figure}
%EndExpansion
%

\textit{F-type}:  The SWDOS\ consists of five vHs as shown in fig. \ref{fig_disp}(a). \ The vHs corresponding to zone boundary spin waves along the $c$-axis or in the $ab$-plane lead directly to the corresponding exchange constants. Other zone boundary spin waves give sums of the exchange constants. Figure \ref{fig_dos2}(a)\ shows how the spectral features of the F-type magnet develop as the ratio $|J_{c}/J_{ab}|$ changes. In the limit that $J_{c}=0$, the SWDOS of a 2-D ferromagnet consists of a sharp peak at $4|J_{ab}|S$ and an upper cutoff at $8|J_{ab}|S$.\

\textit{A-type}: The SWDOS\ consists of several vHs as indicated in fig. \ref{fig_disp}(b). \ The most useful for determining the exchange constants (in units of $\left\vert J_{ab}\right\vert S$) are the maximum
spinwave cutoff [$2(4+|J_{c}/J_{ab}|)$], the high energy edge of the central
band [$2(2+|J_{c}/J_{ab}|)$], and the low energy cusp ($2|J_{c}/J_{ab}|$).
\ Identification of any two of these vHs is sufficient to determine both
$J_{ab}$ and $J_{c}$. \ For example, $J_{ab}$ can be determined by the
difference of the two highest energy vHs identified here (equal to
$2|J_{ab}|S$). \ Figure \ref{fig_dos2}(b)\ shows how the spectral features of
the A-type antiferromagnet develop as the ratio $|J_{c}/J_{ab}|$ changes. \ As
$J_{c}$ becomes relatively small (in the approach to a 2D ferromagnet), the
central band collapses to a single energy at $4|J_{ab}|S$, the cusp just below
the cutoff energy disappears, and the lowest energy cusp at $2J_{c}S$
disappears. This is of course identical to the same limit in the F-type magnet
discussed above.

\textit{C-type}: The SWDOS\ consists of three vHs, as shown in fig. \ref{fig_disp}(c). \ The lowest energy peak\ is due to zone boundary spin waves in the basal plane and depends only on $J_{ab}S$, while the splitting of the two main peaks in the SWDOS gives $J_{c}S$ directly. \ The development of the spectral features of the C-type antiferromagnet with the ratio $|J_{c}/J_{ab}|$ are shown in fig.\ref{fig_dos2}(c). \ As $J_{c}$ becomes relatively small in the limit of the 2-D antiferromagnet, the two main peaks merge into a single peak at $4J_{ab}S$.

\textit{G-type}: The SWDOS\ consists of a single sharp peak at the cutoff energy with a very weak cusp-like vHs just below the cutoff energy. \ The energy of the peak in the DOS is determined by the average exchange $\left\langle J\right\rangle =\left( 4J_{ab}+2J_{c}\right)/6$. \ As shown in fig. \ref{fig_dos2}(d), varying the ratio $J_{c}/J_{ab}$ shifts the entire spectrum. \ The weak second vHs is often masked by the finite energy resolution of the neutron spectrometer and in this case it is possible to determine only the average exchange $<J>$.

\section{Measurements}

In order to test the predictions made in the Heisenberg model calculations
above, inelastic neutron scattering (INS) measurements were performed on the
Pharos spectrometer at the Lujan Center of Los Alamos National Laboratory.
\ Pharos is a direct geometry time-of-flight spectrometer and measures the
scattered intensity over a wide range of energy transfers ($\hbar\omega$) and
angles between 1$^{o}$ - 140$^{o}$ allowing determination of a large swath of
the scattered intensity, $S(Q,\omega)$, as a function of momentum transfer
($\hbar Q$) and $\hbar\omega$.

%TCIMACRO{\TeXButton{B}{\begin{table}[tbp] \centering}}%
%BeginExpansion
\begin{table}[b] \centering
%EndExpansion
\caption{Experimental conditions and data analysis parameters for neutron scattering measurements on LaMnO$_3$, LaVO$_3$, and LaFeO$_3$}%
\begin{tabular}
[c]{ccccc}\hline\hline
Compound & magnetic & $E_{i}$ (meV) & low angle & high angle\\
& ordering &  & range & range\\\hline
LaMnO$_{3}$ & A-type & 75 & 12-42$^{o}$ & 60-120$^{o}$\\
LaVO$_{3}$ & C-type & 75 & 7-32$^{o}$ & 60-110$^{o}$\\
LaFeO$_{3}$ & G-type & 160 & 1-31$^{o}$ & 55-95$^{o}$\\\hline\hline
\end{tabular}
\label{neutrontable}%
%TCIMACRO{\TeXButton{E}{\end{table}}}%
%BeginExpansion
\end{table}%
%EndExpansion
%
Powder samples of LaMnO$_{3}$ (LMO), LaVO$_{3}$ (LVO), and LaFeO$_{3}$ (LFO)
were prepared by conventional solid-state reaction method and subsequently
annealed to tune oxygen stoichiometry. Samples weighed approximately 50 grams
each and were characterized for phase purity by x-ray powder diffraction.
Powders were packed in flat aluminum cans oriented at 45$^{o}$ or 135$^{o}$ to
the incident neutron beam and INS spectra for LMO, LVO, and LFO, were measured
with incident energies ($E_{i}$) of 75, 75, and 160 meV, respectively. The
time-of-flight data were reduced into $\hbar\omega$ and scattering angle
($2\theta$) histograms and corrections for detector efficiencies, empty can
scattering, and instrumental background were performed.\

%TCIMACRO{\FRAME{ftbpFU}{3in}{2.0003in}{0pt}{\Qcb{(color online) (a) Inelastic
%neutron scattering intensity of LaFeO$_{3}$ (color scale) versus scattering
%angle and energy transfer at $T=10$ K and $E_{i}=160$ meV. Horizontal white
%lines delineate regions where phonon and magnetic scattering are isolated. (b)
%Neutron intensity summed over the angle range from 55 - 95$^{o}$ originating
%from phonons. (c) Neutron intensity summed over the low angle range from 1 -
%30$^{o}$ (dots) and phonon background from scaled from high angle sum (magenta
%hatched region) (d) Isolated magnetic scattering from LFO (green) and LSFO
%(red) at $T=$ 10 K.}}{\Qlb{fig_subt}}{fig4.eps}%
%maintain-aspect-ratio TRUE;  display "USEDEF";  valid_file "F";  width 3in;
%height 2.0003in;  depth 0pt;  original-width 0pt;  original-height 0pt;
%cropleft "0";  croptop "1";  cropright "1";  cropbottom "0";
%filename 'fig4.eps';file-properties "XNPEU";}} }%
%BeginExpansion
\begin{figure}
[ptb]
\includegraphics[
scale=0.85
]%
{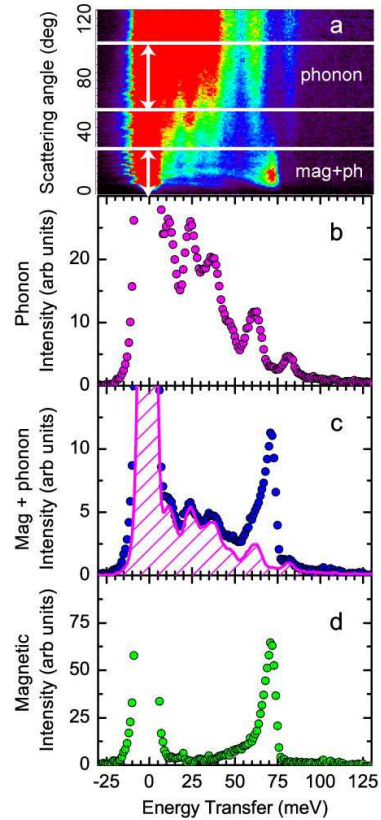}%
\caption{(color online) (a) Inelastic neutron scattering intensity of
LaFeO$_{3}$ (color scale) versus scattering angle and energy transfer at
$T=10$ K and $E_{i}=160$ meV. Horizontal white lines delineate regions where
phonon and magnetic scattering are isolated. (b) Neutron intensity summed over
the angle range from 55 - 95$^{o}$ originating from phonons. (c) Neutron
intensity summed over the low angle range from 1 - 30$^{o}$ (dots) and phonon
background from scaled from high angle sum (magenta hatched region) (d)
Isolated magnetic scattering from LFO (green) at $T=$ 10 K.}%
\label{fig_subt}%
\end{figure}
%EndExpansion
%
Unpolarized inelastic neutron scattering contains contributions from both
magnetic and phonon scattering. \ In order to isolate the spin wave spectrum,
the magnetic scattering must be separated from the phonon scattering. \ This
is accomplished by using the fact that the magnetic scattering falls off with
$Q$ (or $2\theta$) due to the magnetic form factor, while phonon scattering
increases like $Q^{2}$. Figure \ref{fig_subt}(a) shows the full spectrum for
LFO at $T=10$ K as a function of $2\theta$ and $\hbar\omega$, as reported previously.\cite{mcq2007} The band at 75
meV has intensity that falls off with $2\theta$, indicating that it is
magnetic in origin. \ Data summed over the high angle range contains only
phonon scattering (fig. \ref{fig_subt}(b)), while the low angle range contains scattering from both phonons and spin waves (fig.\ \ref{fig_subt}(c)). The magnetic scattering component can be obtained by subtracting the high angle data from low angle data after scaling by a constant factor, as shown in fig.\ \ref{fig_subt}(c) for LFO. \ Figure \ref{fig_subt}(d) shows that the resulting magnetic intensity
for LFO indeed consists of a single peak at $\hbar\omega\sim75$ meV consistent
with the G-type SWDOS shown in fig.\ \ref{fig_dos2}(d).

The strong peak at 0 meV is elastic scattering and very weak peaks at $\sim$20 and 30 meV arise from
imperfect phonon subtraction. The subtraction of the phonon intensity is
subjected to error, primarily due to the fact that the scaling of phonon
intensity from high angles to low angles is only expected to work for incoherent scattering from a monatomic sample.  For real multicomponent samples, the phonon intensity may not scale uniformly to low-$Q$ due to coherent scattering effects (dependence of the phonon cross-section on $Q$) and also due to the different Debye-Waller factors for each component. It is difficult to quantify this error without detailed phonon models, however, based on the general agreement between the isolated magnetic scattering and the calculations discussed below, the introduced errors are often small.

%TCIMACRO{\FRAME{ftbpFU}{2.0435in}{4.1208in}{0pt}{\Qcb{(color online) Extracted
%angle-averaged magnetic intensity (dots) versus the intensity calculated from
%a Heisenberg model for the spin waves for (a)\ LaMnO$_{3}$, (b)\ LaVO$_{3}$,
%and (c) LaFeO$_{3}$.}}{\Qlb{fig_compare}}{fig5.eps}%
%width 2.0435in;  height 4.1208in;  depth 0pt;  original-width 3.1886in;
%original-height 6.4723in;  cropleft "0";  croptop "1";  cropright "1";
%cropbottom "0";  filename 'fig5.eps';file-properties "XNPEU";}} }%
%BeginExpansion
\begin{figure}
[ptb]
\includegraphics[
scale=0.85
]%
{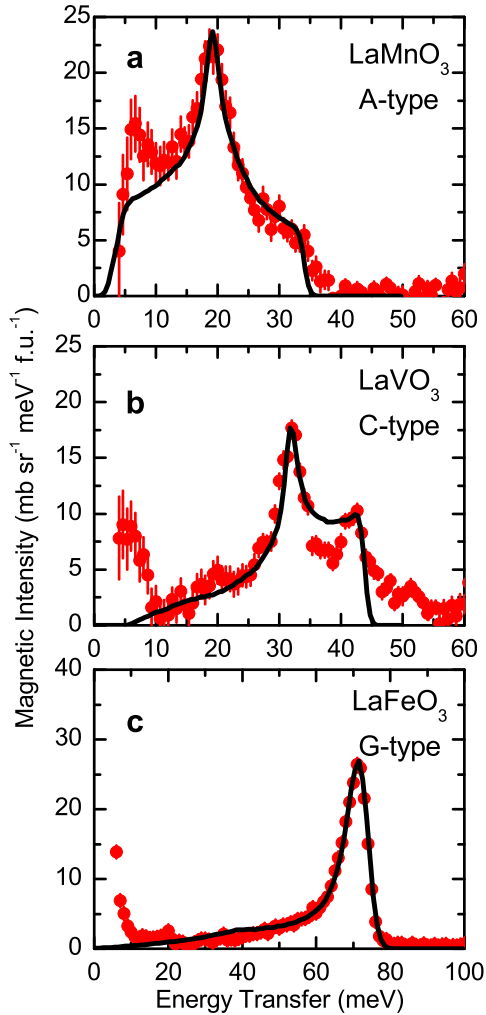}%
\caption{(color online) Extracted angle-averaged magnetic intensity (dots)
versus the intensity calculated from a Heisenberg model for the spin waves for
(a)\ LaMnO$_{3}$, (b)\ LaVO$_{3}$, and (c) LaFeO$_{3}$.}%
\label{fig_compare}%
\end{figure}
%EndExpansion

\ The LMO and LVO data were treated in a similar fashion to the LFO data.
\ Figure \ref{fig_compare} shows the isolated magnetic intensity for the three
different antiferromagnets. In each case, the magnetic spectra share similar features to the respective calculated SWDOS shown in figs.\ \ref{fig_dos1} and \ref{fig_dos2}.

\section{Calculations of the scattered intensity}

\begin{figure}
[ptb]
\includegraphics[
scale=0.6
]%
{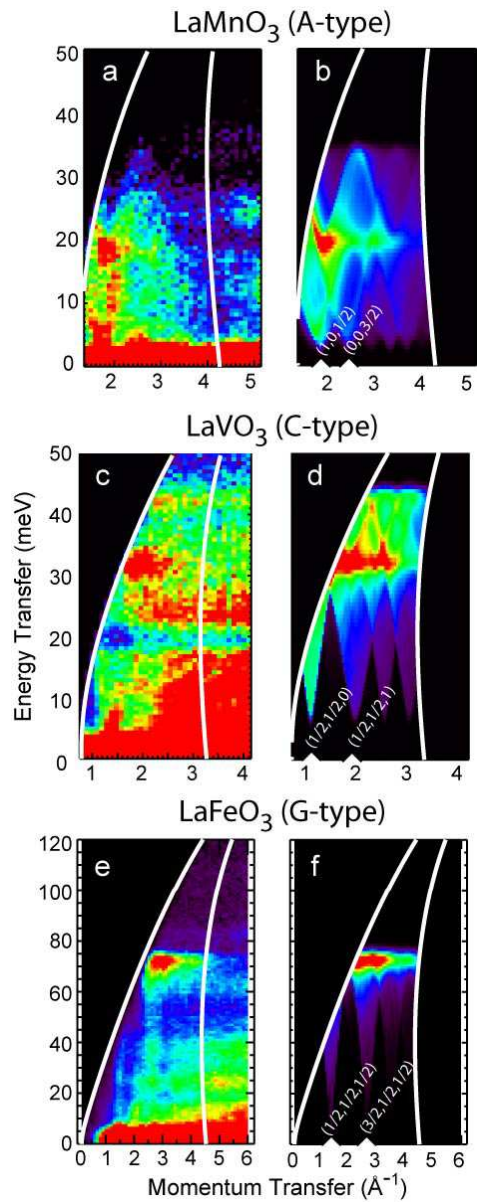}%
\caption{Panels show the measured (left) and calculated (right) neutron
intensities for (a)-(b) A-type LaMnO$_{3}$, (c)-(d) C-type LaVO$_{3}$, and
(e)-(f) G-type LaFeO$_{3}$. \ Experimental conditions and calculation
parameters are given in the text. \ For the measured data in panels (a), (c),
and (e), phonon intensities have not been subtracted and lead to a more
complicated spectral image as discussed in the text. \ For each panel, the
curved white lines indicate the low-angle summation regions leading to the
corrected magnetic spectra in fig. 5.}%
\label{fig_sqw}%
\end{figure}

When performing an INS experiment on a powder, the resulting INS intensities
arise from the averaging of the inelastic scattering structure factor
$S(\mathbf{Q},\omega)$ over all orientations of the crystallites. \ Despite
the orientational averaging, the spectra can show evidence of the spin wave
dispersions, especially at low angles (within the first Brillouin zone) and in the vicinity of the first few magnetic Bragg peaks. \ Such dispersive features are clearly seen in the intensity plots of
$S(Q,\omega)$ in figs \ \ref{fig_sqw}(a), (c), and (e)\ for LMO, LVO, and LFO,
respectively. \ Due to the weighting of the spin wave modes by coherent
scattering intensities, the $Q$-averaged intensity, $S(\omega)$, as shown in
fig. \ref{fig_compare} does not necessarily give the SWDOS. \ This is only
true in the incoherent scattering approximation, which does not apply to the
case of scattering from a magnetically ordered system. \ Therefore, model calculations of the powder averaged spin wave intensities are necessary for accurate comparison to the data.

Numerical calculations of the spin waves in the linear approximation to the
Heisenberg model give not only the dispersion relation $\omega_{n}%
(\mathbf{q})$ for the $n^{th}$ (degenerate) branch (as shown in eqn.\ \ref{eqn1}), but also the spin wave eigenvectors, $T_{ni}(\mathbf{q})$, for the $i^{th}$ spin in the magnetic unit cell. \ The dispersion and associated eigenvectors can be used to calculate the spin wave structure factor for unpolarized neutron energy loss scattering from a single-crystal sample, $S_{mag}(\mathbf{Q},\omega)$.
\begin{equation}
\begin{split}
S_{mag}(\mathbf{Q},\omega) & = \frac{1}{2}\left(  \gamma r_{o}\right)^{2} \left( 1+\frac{(\mathbf{\hat{\mu}\cdot Q)}^{2}}{Q^{2}} \right)\\
& \times\sum\limits_{n}\left\vert \sum\limits_{i}F_{i}(\mathbf{Q})\sigma
_{i}\sqrt{S_{i}}T_{ni}(\mathbf{q})e^{-i\mathbf{Q}\cdot\mathbf{d}_{i}%
}\right\vert ^{2}\\
& \times(n(\omega)+1)\delta(\omega-\omega_{n}(\mathbf{q}))
\end{split}
\end{equation}
where the $i^{th}$ spin with magnitude $S_{i}$ pointed in direction
$\mathbf{\hat{\mu}}$ is located at position $\mathbf{d}_{i}$. \ $\sigma
_{i}=\pm1$ is the direction of the spin relative to the quantization axis
$\mathbf{\hat{\mu}}$ for a collinear spin structure. \ $\mathbf{q}%
=\mathbf{Q}-\mathbf{\tau}$ is the spin wave wavevector in the first Brillouin
zone. \ Finally, the function $n\left(  \omega\right)  $ is the temperature
dependent Bose factor and $F_{i}(\mathbf{Q})=\frac{1}{2}g_{i}f_{i}%
(\mathbf{Q})e^{-W_{i}(\mathbf{Q})}$ is a product of the Lande $g$-factor,
magnetic form factor, and Debye-Waller factor for the $i^{th}$ spin,
respectively. \ The constant $\left(  \gamma r_{o}\right)  ^{2}=$ 290.6 millibarns allows calculations of the cross-section to be reported in absolute units of (millibarns Steradian$^{-1}$ meV$^{-1}$ (formula~unit)$^{-1}$).For the simple perovskite magnets studied here, all ions in the magnetic cell are considered to be equivalent. \ The structure factor can then be written,

\begin{equation}
\begin{split}
S_{mag}(\mathbf{Q},\omega) & =\frac{1}{2}\left(  \gamma r_{o}\right)^{2} SF^{2}(\mathbf{Q})\left( 1+\frac
{(\mathbf{\hat{\mu}\cdot Q)}^{2}}{Q^{2}} \right)\\
& \times\sum\limits_{n}\left\vert \sum\limits_{i}\sigma_{i}T_{ni}%
(\mathbf{q})e^{-i\mathbf{Q}\cdot\mathbf{d}_{i}}\right\vert ^{2}\\
& \times(n(\omega)+1)\delta(\omega-\omega_{n}(\mathbf{q}))
\end{split}
\label{eqn5}
\end{equation}

In the calculations below, we use the isotropic magnetic form factors found in
the International Crystallography Tables \cite{magformfactor} and the
Debye-Waller factor is set equal to one. \ The differential magnetic
cross-section is proportional to the structure factor.
%
%TCIMACRO{\FRAME{ftbpFU}{3in}{2.0003in}{0pt}{\Qcb{Panels show the measured
%(left) and calculated (right) neutron intensities for (a)-(b) A-type
%LaMnO$_{3}$, (c)-(d) C-type LaVO$_{3}$, and (e)-(f) G-type LaFeO$_{3}$.
%\ Experimental conditions and calculation parameters are given in the text.
%\ For the measured data in panels (a), (c), and (e), phonon intensities have
%not been subtracted and lead to a more complicated spectral image as discussed
%in the text. \ For each panel, the curved white lines indicate the low-angle
%summation regions leading to the corrected magnetic spectra in fig. 5.}%
%}{\Qlb{fig_sqw}}{fig6.eps}{\special{ language "Scientific Word";
%type "GRAPHIC";  maintain-aspect-ratio TRUE;  display "USEDEF";
%valid_file "F";  width 3in;  height 2.0003in;  depth 0pt;
%original-width 0pt;  original-height 0pt;  cropleft "0";  croptop "1";
%cropright "1";  cropbottom "0";  filename 'fig6.eps';file-properties "XNPEU";}%
%} }%
%BeginExpansion
%EndExpansion

%TCIMACRO{\TeXButton{B}{\begin{table}[tbp] \centering}}%
%BeginExpansion
\begin{table}[tbp] \centering
%EndExpansion
\caption{Values of the ionic spin, exchange energies, and uniaxial anisotropy energies used in calculations of the neutron scattering intensity from spin waves.}%
\begin{tabular}
[c]{ccccc}\hline\hline
Compound & $S$ & $J_{ab}$ (meV) & $J_{c}$ (meV) & $g\mu_{B}H_{a}$
(meV)\\\hline
LaMnO$_{3}$ & 2 & 1.85 & -1.1 & 0.6\\
LaVO$_{3}$ & 1 & -7.8 & 2.9 & 0.6\\
LaFeO$_{3}$ & 5/2 & -4.87 & -4.87 & 0\\\hline\hline
\end{tabular}
\label{exchange}%
%TCIMACRO{\TeXButton{E}{\end{table}}}%
%BeginExpansion
\end{table}%
%EndExpansion

To compare Heisenberg model spin wave results to the powder INS data,
powder-averaging of $S_{mag}(\mathbf{Q},\omega)$ is performed by Monte-Carlo
integration over a large number of $\mathbf{Q}$-vectors lying on a
constant-$Q$ sphere, giving the orientationally averaged $S_{mag}(Q,\omega)$
which depends only on the magnitude of $Q$. Figs \ref{fig_sqw}(b),(d), and (f) show
calculations of $S_{mag}(Q,\omega)$ (broadened by instrumental resolution) at
10 K for LMO, LVO, and LFO, respectively, and can be compared to the
corresponding data in figs. \ref{fig_sqw}(a), (c), and (e). In order to properly
calculate the structure factor, we use the structural parameters for
the three compounds (which are orthorhombically distorted perovskite
structures). \ However, only the cubic exchange interactions $J_{ab}$
and $J_{c}$ are employed. \ The values for $J_{ab}$, $J_{c}$, and $H_{a}$ used
in the calculations for each compound are shown in Table \ref{exchange}.\ In
the case of LMO, the anisotropy field was determined by Hirota \textit{et al.}
from single-crystal dispersion measurements.\cite{hirota1996}, and the
exchange constants can be compared to the values obtained in that paper. \ For
LVO, the anisotropy field was determined from cold neutron measurements of the
anisotropy gap in powder samples.\cite{yan_unpub} The small anistropy energies reported
here have very little effect on the energy of the zone boundary spin waves, which are determined
primarily by the exchange.

\ The calculations can be summed over scattering angles in order to compare the equivalent angle-summed data, as shown in fig \ref{fig_compare}. Overall, the agreement between the data and calculations is excellent. \ This is a
testament to the effectiveness of the Heisenberg model for these compounds in predicting not only the spin wave energies, but also the intensities. \ However, some differences observed in the comparison of data and
calculation are worth noting. \ At low energies near to the elastic line, additional intensity is observed, most notably in the 5 - 10 meV range in $E_{i}=$75 meV data. The origin of this intensity is unclear, but it is possible that it originates from multiple elastic scattering.  For LaMnO$_{3}$, this additional scattered intensity, combined with insufficient elastic energy resolution, does not allow the observation of the low energy vHs expected at $\sim$6 meV that can be used to determine $J_{c}$. \ Higher resolution measurements are required to obtain $J_{c}$ exclusively.

The poorest agreement between data and calculation occurs for LaVO$_{3}$ in fig \ref{fig_compare}(b). \ While the data shows clear vHs at $\sim$32 and $\sim$44 meV, the calculated intensity shows a shoulder, rather than a clear peak in the upper vHs, indicating that the Heisenberg model does not reproduce the LVO spin wave intensities with the same accuracy as for LMO and LFO. \ Much of this discrepancy may be due to the physics of LVO, where C-type magnetic strucure can arise from either weak Jahn-Teller driven orbital ordering or orbital singlet formation.  Competition between these two scenarios require spin-orbital coupling terms that go beyond the Heisenberg Hamiltonian.\cite{khaliullin2001,ulrich2003} \ Also, additional intensity at 20 meV and 50 meV in LVO is likely due to improper phonon subtraction. \ This subtraction is made more difficult due to the small spin ($S=1$) of the V ion, which leads to weaker spin wave scattering (see eqn. \ref{eqn5}). \

Figure \ref{fig_sqw} shows a comparison of the measured and calculated scattered intensities as a function of $Q$ and $\hbar\omega$ for the three compounds. The calculation results in figs.\ \ref{fig_sqw}(b), (d), and (f) show clearly the coherent scattering of the powder averaged spin waves. \ The most obvious coherent scattering feature is the necking down of acoustic spin waves in the vicinity of allowed magnetic Bragg reflections.\ The characteristic ordering wavevectors for the different antiferromagnets are; (0,0,1/2) for A-type, (1/2,1/2,0) for C-type, and (1/2,1/2,1/2) for G-type (using the cubic indexing). The first two observed magnetic Bragg peaks in each case are indicated\ in fig. \ref{fig_sqw}. \ Additional coherent scattering features can also be seen for zone boundary spin waves, where intensities tend to peak in between the allowed magnetic Bragg peaks. \ Fig. \ref{fig_sqw} enforces the general agreement of the Heisenberg model calculations of the spin wave intensity with neutron scattering measurements.
\ Unfortunately, the comparison of the Heisenberg model spin wave intensities to the data is complicated because measurements also contain coherent phonon scattering intensity. \ The phonon intensity bands present themselves mainly as horizontal (constant energy) streaks.  A prominent phonon band can be seen, for example, at 25 meV in LVO (fig \ref{fig_sqw}(c)), and at 25, 40, and 60 meV in LFO ((fig \ref{fig_sqw}(e)).
%
%TCIMACRO{\FRAME{ftbpFU}{3.7948in}{3.2033in}{0pt}{\Qcb{(color online)\ The
%Q-dependence of the neutron scattering data for different energy transfer
%ranges in LaMnO$_{3}$: (a)\ 8-12 meV, (b) 12-16 meV, (c)\ 18-22 meV, and
%(d)\ 26-30 meV. \ The red dots are the experimental data. The dashed line it
%an estimate of the incoherent phonon background plus multiple scattering. The
%solid line is the calculation of the polycrystalline averaged spin wave
%scattering plus background using the parameters in the text.}}{\Qlb{fig_Qdep}%
%}{fig7.eps}{\special{ language "Scientific Word";  type "GRAPHIC";
%width 3.7948in;  height 3.2033in;  depth 0pt;  original-width 3.7464in;
%original-height 3.1583in;  cropleft "0";  croptop "1";  cropright "1";
%cropbottom "0";  filename 'fig7.eps';file-properties "XNPEU";}} }%
%BeginExpansion
\begin{figure}
[ptb]
\includegraphics[
scale=0.9
]%
{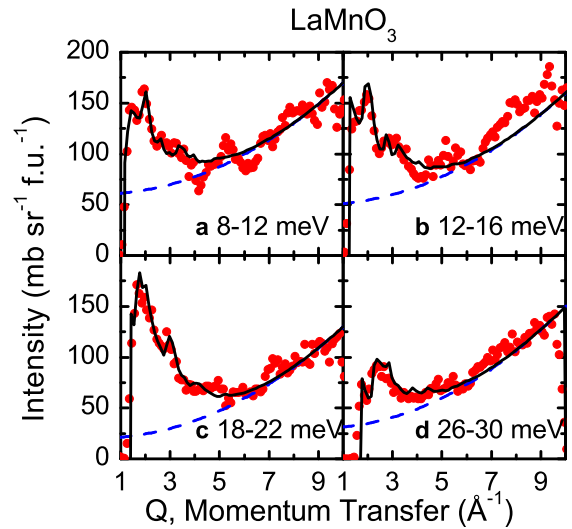}%
\caption{(color online)\ The Q-dependence of the neutron scattering data for
different energy transfer ranges in LaMnO$_{3}$: (a)\ 8-12 meV, (b) 12-16 meV,
(c)\ 18-22 meV, and (d)\ 26-30 meV. \ The red dots are the experimental data.
The dashed lines are estimates of the incoherent phonon background plus
multiple scattering. Solid lines are calculations of the polycrystalline
averaged spin wave scattering using the parameters in the text plus background.}%
\label{fig_Qdep}%
\end{figure}
%EndExpansion

The success of the Heisenberg model in estimating the measured spin wave intensities is better observed by plotting constant energy $Q$-cuts, as shown in fig. \ref{fig_Qdep} for LMO. \ The plots show $Q$-oscillations of the experimental magnetic spin wave scattering above a background due mainly to phonon scattering and background/multiple scattering.\ A constant background and incoherent phonon scattering intensity (proportional to $Q^{2}$) are added to the calculated spin wave scattering in order to compare to the measured data. \ The agreement is excellent. \ The overall consensus is that the spin wave intensities are well represented by the Heisenberg model and it is promising that one can obtain more from powder data than just an estimate of the spin wave DOS.\ Analysis of the full structure factor $S_{mag}(Q,\omega)$ may allow exchange interactions to be determined in more complicated magnetic structures, or with interactions beyond nearest-neighbor. However, a full analysis of powder averaged spin waves requires better understanding of the phonon spectra and multiple scattering. \ In the future, we plan on combining fully coherent calculations of both phonons and spin waves to attempt a more ambitious
analysis of the full $S(Q,\omega)=S_{mag}(Q,\omega)+S_{phonon}(Q,\omega)$.\cite{mcq_unpublished}

\section{Summary}

We have demonstrated that inelastic neutron scattering experiments on powders,
in combination with calculations of the spin wave scattering in a Heisenberg
model can give detailed information about the exchange interactions in simple
magnets. \ The agreement of not only the spin wave DOS, but also the
$Q$-dependence of coherent features in the spin wave scattering gives hope that
even more complicated magnetic systems can be analyzed using the full
$S(Q,\omega)$. \ The advent of new spallation neutron sources, such as the
Spallation Neutron Source, will allow the rapid measurements of samples with
good statistics and make detailed systematic studies of magnetism possible.

\appendix{}

\begin{center}
ACKNOWLEDGMENTS
\end{center}

RJM would like to thank F.\ Trouw, A.\ Llobet, and M.\ Hehlen for assistance with Pharos. \ Ames Laboratory is supported by the U. S.
Department of Energy Office of Science under Contract No. DE-AC02-07CH11358.
The work has benefited from the use of the Los Alamos Neutron Science Center
at Los Alamos National Laboratory. LANSCE is funded by the U.S. Department of
Energy under Contract No.W-7405-ENG-36.

\bibliographystyle{apsrev}
\bibliography{acompat,spin_wave_DOS_prb}

\end{document}